\documentclass[runningheads]{llncs}
\usepackage{graphicx}
\usepackage{url}
\begin{document}
\title{VADIS -- a VAriable Detection, Interlinking and Summarization system}
%
%
\author{Yavuz Selim Kartal\inst{1}\orcidID{0000-0002-2146-2680} \and
Muhammad Ahsan Shahid\inst{1}\orcidID{0000-0002-7274-7934} \and
Sotaro Takeshita\inst{2, 3}\orcidID{0000-0002-6510-7058} \and
Tornike Tsereteli\inst{2}\orcidID{0000-0003-4298-3570} \and
Andrea Zielinski\inst{4}\orcidID{0000-0003-2092-6153} \and
Benjamin Zapilko\inst{1}\orcidID{0000-0001-9495-040X} \and
Philipp Mayr\inst{1}\orcidID{0000-0002-6656-1658}}
\authorrunning{Y. Kartal et al.}
%
\institute{GESIS -- Leibniz Institute for the Social Sciences, Cologne, Germany 
\email{firstname.lastname@gesis.org} \and
University of Mannheim, Germany \\
\email{{sotaro.takeshita, tornike.tsereteli}@uni-mannheim.de} \and
Hochschule Mannheim, Germany \and
Fraunhofer ISI, Karlsruhe, Germany \\
\email{andrea.zielinski@isi.fraunhofer.de}
} 
\maketitle              
\begin{abstract}
The VADIS system addresses the demand of providing enhanced information access in the domain of the social sciences. This is achieved by allowing users to search and use survey variables in context of their underlying research data and scholarly publications which have been interlinked with each other.


\end{abstract}
\section{
Introduction}

In the social sciences, as in other scientific disciplines, there is a growing necessity to support researchers by providing enhanced information access through linking of publications and underlying datasets. One of the key artifacts in social science studies are concepts, namely survey variables (SV), which have been used to measure sociological phenomena. These variables are generally the access to research data \cite{friedrich2016}, but are neither represented nor interlinked in the information sources relevant for social science research, i.e., scholarly publications.

The key objective of the VADIS (VAriable Detection, Interlinking and Summarization) project \cite{kartal2022} is to allow for searching and using survey variables in context and thereby enhance information access of scholarly publications and help to increase the reproducibility of research results. We combine text mining techniques and semantic web technologies that identify and exploit links between publications, their topics, and the speciﬁc variables that are covered in the surveys.
These semantic links build the basis for the VADIS system which offers users better access to the scientific literature by, e.g., search, summarization, and linking of identified survey variables.

In detail, the system follows three objectives: (1) serving as a demonstrator for the summarization and SV identification methods, (2) enabling search and exploration functionalities addressed in the use cases \cite{kartal2022}, and (3) serving as an evaluation test bed for user studies.

\section{Related Works}

Many systems have been developed concerning the main objectives of VADIS independently. For summarization, there is a line of work focusing on scientific document summarization from domain-specific datasets \cite{Cachola2020TLDRES} to the models that exploit the unique properties such as citation contexts \cite{cohan_scientific_2018} or discourse structures \cite{cohan-etal-2018-discourse,akiyama2021hie}. Semantic Scholar and 
IBM Science Summarizer \cite{erera-etal-2019-summarization} provide extreme and extended summaries, respectively, coupled with search engines in their systems.
There are various search systems in the Social Sciences. 
The UK Data Service, 
The Inter-university Consortium for Political and Social Research (ICPSR), 
the Understanding Society \cite{buck2012understanding}, the Norwegian centre for research data (NSD) 
and GSS Data Explorer 
serve as variable search systems on their specific research datasets. Google Dataset enables dataset search from a wide range of areas, and GESIS Search\footnote{\url{https://search.gesis.org/} \label{url:gesis-search}}   provides an advanced search system for publications, datasets, and their variables in an integrated way.

\begin{figure}
  \centering
      \includegraphics[width=\linewidth,page=2]{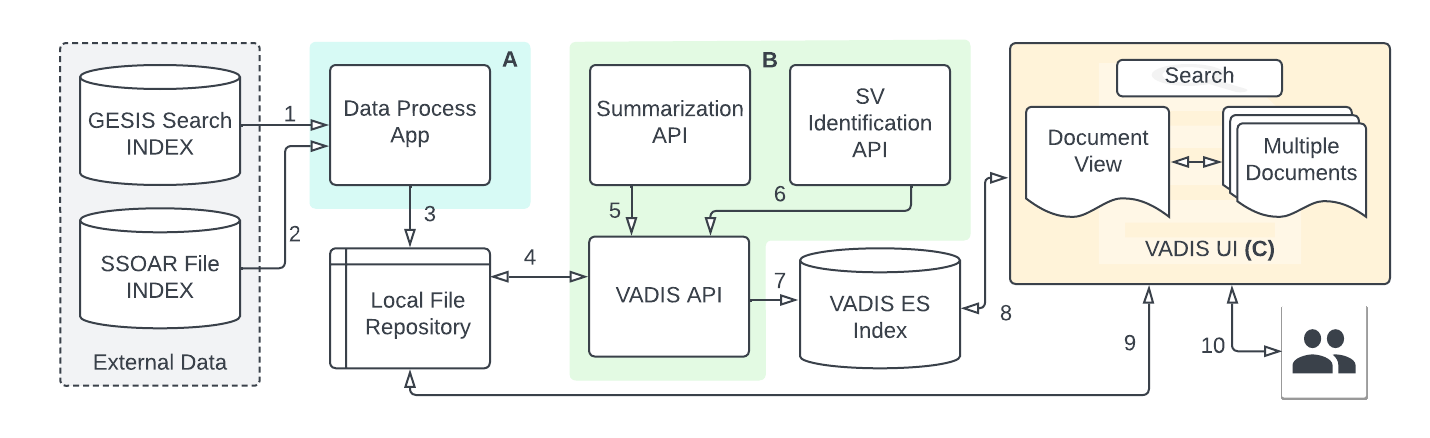}%

      \caption
        {%
          VADIS Framework%
          \label{fig:frame}%
        }%
        \vspace{-10mm}
\end{figure}

\section{System Overview}
The current version of the VADIS system\footnote{\url{https://demo-vadis.gesis.org/}\label{url:vadis}} contains 607 documents in English and German, which are all of the publications with digitally available research dataset linkages in GESIS Search, their extreme summaries, 7086 linkages of their sentences to variables, and their original files.  

We developed a framework that integrates summarization and survey variable identification processes asynchronously and delivers their outcomes to the users. This section explains the VADIS framework consisting of three main components: Data Preprocessing (A), VADIS Modules (B), VADIS UI (C), as shown in Figure~\ref{fig:frame}. All published modules are available via the VADIS Github project: \url{https://github.com/vadis-project/}.

\subsection{Data Preprocessing (A)}
We first retrieve the metadata of the publications and their PDF files from
GESIS Search and SSOAR indices respectively (1, 2).\footnote{The numbers in parentheses show the steps in Figure 1.} This includes relations between publications and research datasets with detailed variable descriptions and the abstract. All data is stored and updated  on a local server (3). Grobid is used to extract plain text from the publication PDFs, while full-text data is kept for the VADIS UI search operations and survey variable identification (3).
 

 
\subsection{VADIS Modules (B)} 

The VADIS Modules are a set of APIs and operations that yield the main outcomes of the project from the preprocessed data. 

\textbf{Summarization API:} This API provides two features. First is an extractive summarization which selects one salient sentence as a summary in a given document. Second is an abstractive summarization that generates a short summary of a given document. The API accepts English and German documents as inputs and generates summaries in English or German. This is achieved by training models on a cross-lingual summarization dataset \cite{takeshita2022x}. We also provide distilled models for a faster and memory-efficient inference that enables generating 50\% more summaries per second without critical quality degradation \cite{takeshita2023cross}.

\textbf{SV Identification API:}  This API has two main functions. The first classifies whether a sentence primarily defines a variable by using an ensemble of a model, fine-tuned on an extension of a previously published dataset \cite{tsereteli-etal-2022-overview}, and a sentence-transformers-based \cite{reimers-2019-sentence-bert} retriever, which syntactically and semantically matches the input to similar texts from the database of variables. Given the classified sentences, the second function then uses the retriever to match a subset of variables filtered by using the publication-research dataset relation information. In combination, the functions present users with a narrow list of possible sentences defining used variables as well as a list of candidate variables.

\textbf{VADIS API:} This is an intermediate API that integrates other APIs, input and output data, and index. As shown in Figure ~\ref{fig:frame}, it provides abstracts to Summarization API (5) and full-text and survey variables to SV Identification API (6) and it fetches their outcomes and inserts them into Elastic Search (ES) index (7). It also inserts metadata of the publications to ES for the search feature.

\subsection{VADIS User Interface (C)}
The UI enables users to interact and explore the novel features available within the responsive and user-friendly demonstrator\textsuperscript{\ref{url:vadis}}(10). 
The VADIS UI directly communicates with the ES index and local file repository containing PDF documents to retrieve data (8, 9). As shown in Figure 2, it provides metadata, extreme summaries, and variable linkages for publications, allowing users to perform classical document search via natural language queries within the extracted full-text. Additionally, it allows users to locate extracted survey variables in the actual PDF,  so that variables can be inspected in context.



\begin{figure}
  \centering
      \includegraphics[width=\linewidth,page=2]{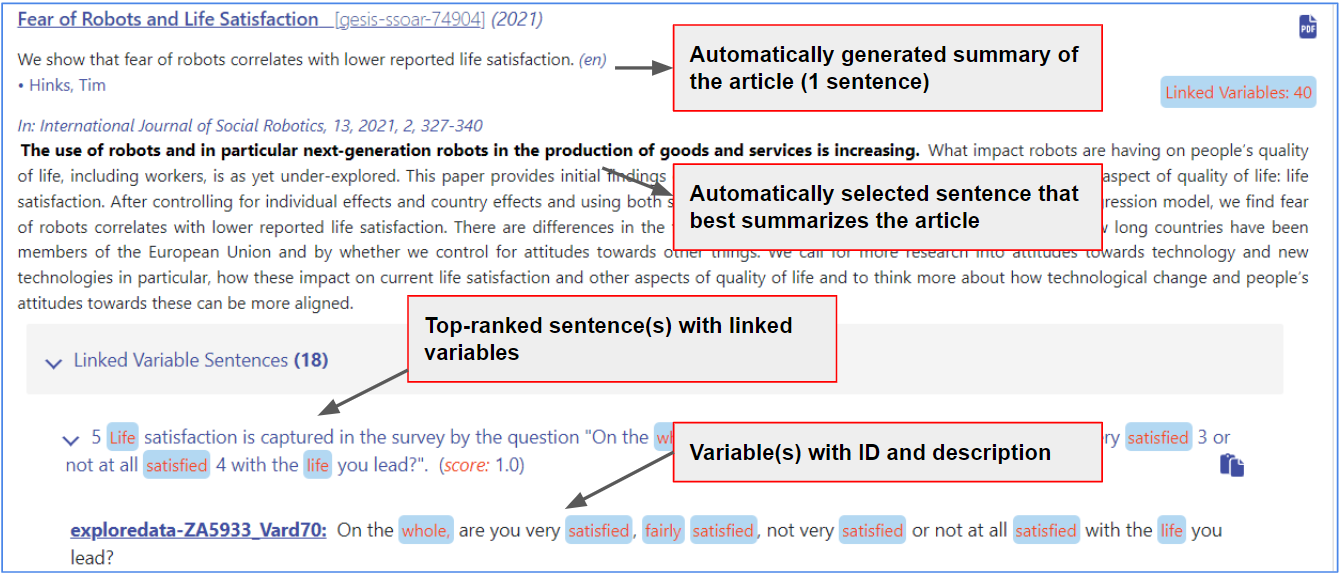}%
      \caption
        {%
          VADIS User Interface%
          \label{fig:ui}%
        }%
        \vspace{-5mm}
\end{figure}

\section{User Test \& Conclusions}
We conducted a preliminary user study with six researchers (5 social science research associates and a student; all postgraduates and experienced users in IR). The user study consisted of two search scenarios to evaluate the usability of the interface and the quality of the search results. During the tests, the users were asked to \textit{think out loud} and in the end they should answer questions related to the search experience. Details of the search scenarios and the questions are available via the following repository:  \url{https://github.com/vadis-project/user-tests} .\\
\textbf{Finding variables and datasets}. When searching for publications by keywords, users found useful links to survey variables (including the associated metadata), revealing a new direction to design survey questionnaires, and showing “new perspectives that I would not have found otherwise”. Also, most participants found the \textbf{extreme summarization} along with metadata helpful.\\
\textbf{Limitations}. Users suggested ranking publications by the “number of detected variables”, apart from classical IR measures such as relevancy and recency. In addition to the annotation of variables in the corresponding PDF of the scientific paper, students suggested a better (graphical) display of the linked variables on the result page. Moreover, it was found that the highlighting of common words in scientific documents and variables could be improved by focusing on important social science concepts.



Following a user-centred design and development process in the VADIS project, the results of the user tests give us directions for further improvements of the system.
Eventually, positively evaluated features will perspectively be implemented into the GESIS search and will thus be available in a widely used search system for the social sciences.

\textbf{Acknowledgement:} This work was supported by the DFG project VADIS under grant numbers: ZA 939/5-1, PO 1900/5-1, EC 477/7-1, KR 4895/3-1.

%
%
%
\bibliographystyle{splncs04}
\bibliography{ecir24}

\end{document}